\def \D{{\cal D}}
\def\e{{\epsilon}}

\def\eo{\epsilon_{\rm obs}}

\def\to{t_{\rm obs}}

\def\tauo{\tau_{\rm obs}}
\def\eh{\epsilon_{\rm H}}
\def\E{{\cal E}}
\def\cross{{\tt T}}

\documentstyle[aaspp4]{article}

\tighten
\eqsecnum

\begin{document}

\title{ELECTRON ACCELERATION AND SYNCHROTRON \\ RADIATION IN DECELERATING PLASMOIDS}

\author{Charles D. Dermer and James Chiang}
\affil{E. O. Hulburt Center for Space Research, Code 7653, Naval Research
Laboratory, Washington, DC 20375-5352}

\begin{abstract}

An equation is derived to calculate the dynamics of relativistic magnetized plasma 
which decelerates by sweeping up matter from the ISM. Reduction to the non-radiative and radiative
regimes is demonstrated for the general case of a collimated plasmoid whose area
increases as a power of distance $x$ from the explosion, and in the specific case of a
blast-wave geometry where the area increases quadratically with $x$.  An equation for the evolution of the
electron momentum distribution function in the comoving fluid frame is used to
calculate the observed synchrotron radiation spectrum. The central uncertainty involves the
mechanism for transfering energy from the nonthermal protons to the electrons. The simplest
prescription is to assume that a fixed fraction of the comoving proton power is instantaneously
transformed into a power-law, ``shock"-like electron distribution function. This permits an
analytic solution for the case of a relativistic plasmoid with a constant internal
magnetic field. Effects of parameter changes are presented.
Breaks in the temporal behavior of the primary burst emission and long wavelength afterglows occur
on two time scales for external ISM density distributions $n_{\rm ext} \propto x^{-\eta}$.  The
first, as emphasized by  M\'esz\'aros
\& Rees, represents the time when the outflow sweeps up $\approx {\cal M}_{\rm th}/\Gamma_0$ of
material from the ISM, where ${\cal M}_{\rm th}$ is the mass in the outflow ejecta and $\Gamma_0$
is the initial Lorentz factor of the plasmoid. The second represents the time when synchrotron
cooling begins strongly to regulate the number of nonthemal electrons that are producing radiation
which is observed at a given energy. 

Results are applied to GRB and blazar variability. The slopes of the time profiles of the X-ray and
optical light cuves of the afterglows of GRB 970228 and GRB 970508 are explained by injection of hard
electron spectra with indices $s < 2$ in a deceleration regime near the radiative limit.  
We also propose that blazar flares are due to the transformation of the directed
kinetic energy of the plasmoid when interacting with the external medium, as would occur
if relativistic outflows pass through clouds orbiting the central supermassive black hole.
 
\end{abstract}

\section{Introduction}

{\it CGRO} BATSE and {\it Beppo-SAX} observations of GRBs have renewed interest in relativistic
blast-wave models.  The BATSE observations (Meegan et al.\ \markcite{Meegan92}1992; Fishman \& Meegan
\markcite{Fishman95}1995) indicate that GRBs originate from cosmological distances and therefore
involve impulsive releases of energy exceeding
$10^{51}$ ergs (assuming isotropic emission).  The {\it Beppo-SAX} observations of X-ray
afterglows from GRB 970228 (Costa et al.\ \markcite{Costa97}1997) and GRB 970508 (Piro et al.\
\markcite{Piro97}1997) led to the first optical counterpart identifications of GRB sources (van
Paradijs et al.\ \markcite{vanParadijs97}1997; Djorgovski et al.\ \markcite{Djorgovski97}1997),
culminating in the detection of absorption lines from the optical counterpart to GRB 970508,
showing that it lies at redshift $z \geq 0.835$ (Metzger et al.\  \markcite{Metzger97}1997). The X-ray
emission from GRB 970228 decays as a power-law $\propto
\to^{-1.33\pm 0.12}$ between
$\approx 50$ s and one week following the GRB (Costa et al.\
\markcite{Costa97}1997), where $\to$ is the observer time. The optical counterpart of GRB 970228
decays $\propto \to^{-1.12\pm 0.08}$ between 0.6 days and 189 days following the burst, though it
exhibits apparent fluctuations from a simple power-law (Galama et al.\ \markcite {Galama97}1997; Garcia et
al.\ \markcite{Garcia97}1997).

The X-ray afterglow of GRB 970508 displays a peak at about 2 days following the burst,
with an underlying power law decline $\propto
\to^{-1.1}$ from the end of the main phase of the burst at about 30 sec to $\approx 6.1$ days
following the GRB (Piro et al.\ \markcite{Piro97}1997).  The optical flux of
GRB 970508 appears to be roughly constant between $\approx 6$ hrs and 1 day, whereupon it
subsequently rises and likewise peaks at $\approx 2$ days after the burst (Djorgovski et al.\
\markcite{Djorgovski97}1997; Sahu et al.\ \markcite{Sahu97}1997 and references therein).  The
optical emission thereafter declines $\propto \to^{-1.18\pm 0.04}$ (Garcia et al.\
\markcite{Garcia97}1997).  The 4.86 GHz radio emission from GRB 970508 was
first detected $\approx 6$ days after the GRB, is rapidly variable (evidently due to interstellar
scintillation; Goodman
\markcite{Goodman97}1997) though with an overall flat time profile until $\approx 60$ days
following the burst, after which it begins to monotonically decline (Frail et al.\
\markcite{Frail97}1997). 

Because of these observations, much interest has been focused on the hydrodynamical study of
relativistic blast waves by Blandford \& McKee (\markcite{Blandford76}1976), which has been applied
to GRBs by Rees, M\'esz\'aros, \& coworkers in a series of papers (e.g., M\'esz\'aros \&
Rees \markcite{MR93}1993, \markcite{MR97}1997; M\'esz\'aros,
Rees, \& Papathanassiou \markcite{MRP94}1994; Wijers, Rees, \& M\'esz\'aros
\markcite{WRM97}1997; Panaitescu \& M\'esz\'aros \markcite{Panaitescu97}1997) devoted to
understanding the prompt GRB emision and the long wavelength afterglows.  Vietri
(\markcite{Vietri97a}1997a,\markcite{Vietri97b}b) used the results of Blandford \& McKee in the
radiative regime to explain the delayed X-rays, and Waxman
(\markcite{Waxman97a}1997a,\markcite{Waxman97b}b) applied the blast wave model to the long wavelength
afterglows.  M\'esz\'aros, Rees,
\& Wijers (\markcite{MRW}1997) have derived power-law time profiles for delayed emission in
different regimes assuming analytic models for the ISM density distribution.

Here we reframe the hydrodynamical approach by deriving an equation for plasmoid dynamics in terms
of the comoving particle distribution functions.  This permits the radiation physics to be treated
in the comoving frame, and the plasmoid dynamics to be calculated self-consistently. This
approach can be applied, with appropriate modifications, to temporal studies of 
blazar emissions which are also thought to result from relativistic plasma outflows.  Blazar
flares, which vary on time scales as short as hours to days in the 1 MeV - 1 TeV regime (e.g., Michelson
et al.\ \markcite{Michelson94}1994; Catanese et al.\ \markcite{Catanese97}1997), are in this picture due to
the reconversion of directed kinetic energy of the outflowing plasmoid by interactions with the ISM (see
Hartman et al.\ \markcite{Hartman97}1997 and Shrader \& Wehrle \markcite{Shrader97}1997 for recent
reviews of blazar observations). We suggest that the passage of the relativistic plasma outflow
through a cloud-like density enhancement could initiate such a flare.

In \S 2, we derive an equation for plasmoid dynamics in terms of the evolving comoving particle
distribution functions.  Reduction to well-known forms in the non-radiative and radiative regimes is
performed in \S 3. Prescriptions for the energy transfer from nonthermal protons to electrons, which
get us past the difficult particle acceleration and plasma physics which can be implemented in later
studies, are
proposed in \S 4. In \S 5, we treat the optically thin synchrotron radiation process, and effects of
parameter changes are presented in \S 6. The results are examined analytically in \S 7, and 
application to GRBs and blazars is made in \S 8. We summarize in \S 9. A detailed
numerical treatment is presented in a companion paper (Chiang \& Dermer \markcite{Chiang97b}1997b).

\section{Plasmoid Dynamics}
 
We consider the system shown in Fig.\ 1. A plasmoid, consisting of a volume of magnetized plasma, is
located on the $\hat x$-axis at location $x$ and moves with Lorentz factor $\Gamma(x)$.  The comoving
volume of the plasmoid is $V(x) \cong A(x)L(x)$, where $A(x)$ is the area of the plasmoid at $x$ and
$L(x)$ is the mean width of the plasmoid as measured in the comoving frame. 
 The plasmoid
travels through an external medium with density $n_{\rm ext}(x)$ which 
is assumed to be fully
ionized electron-proton plasma. Consequently the plasmoid captures an electron-proton pair of
the ISM gas if the proton Larmor radius is much smaller than the 
characteristic size scales ($L$
and $A^{1/2}$) of the plasmoid.  Except for chance collisions, neutral particles would pass
through the plasmoid unless they are ionized by interactions with the plasmoid.

The plasmoid is assumed to  consist initially of thermal lepton-proton plasma which was ejected
from the GRB event. In our notation, $m_p = 1.5\times 10^{-3}$ ergs and $m_e = 8.2\times 10^{-7}$
ergs, so that the mass of the thermal particles in the plasmoid is just ${\cal M}_{\rm th} =
m_p N_{\rm th} (1+a_{\rm th})$, where $N_{\rm th}$ is the total number of protons in the
plasmoid ejecta.  The factor $a_{\rm th} = (1+2z_{\rm th})m_e/m_p$ corrects for the mass of
electrons and  pairs, the latter of which are assumed to be present in the ratio $z_{\rm th} =
N_{\rm th,+}/N_{\rm th}$, where $N_{\rm th,+}$ is the number of thermal positrons. Here we assume
that
$k_{\rm B}T_e/m_e \ll 1$ and $k_{\rm B}T_p/m_p \ll 1$, so that the thermal kinetic energy is a
small fraction of the rest mass energy of the particles. 

We furthermore adopt the notation that $N_{p(e)}$ stands for the proton (electron) momentum
distribution function $N_{p(e)}[p;x(t)]$ differential in dimensionless momentum
$p\equiv\beta\gamma$ (not to be confused with the subscript $p$ which stands for proton), and that
$N^\prime_{p(e)}$ stands for the spatial gradient
$\partial N_{p(e)}[p;x(t)]/\partial x$ of the proton (electron) momentum distribution function,
where $t$ is the time in the comoving frame.  If the nonthermal particles are isotropically
distributed in the comoving frame (see discussion in Dermer \& Schlickeiser \markcite{Dermer93}1993),
then the relativistic mass
$\cal M_{\rm nt}$ of nonthermal particles is given by 
$${\cal M}_{\rm nt} = m_p\int_0^\infty dp \;\gamma\; (N_p + a_{\rm nt}N_e)\;,\eqno(1)$$
\noindent where $a_{\rm nt}(p;x) = [1+2z_{\rm nt}(p;x)]m_e/m_p$ is the additional relativistic mass
in the form of nonthermal leptons, and $z_{\rm nt}(p)$ is the momentum-dependent pair ratio in the
nonthermal lepton distribution.

The x-component of the relativistic momentum of the plasmoid at location $x$, denoted $\Pi_x(x)$, is given
by $P({\cal M}_{\rm th}+{\cal M}_{\rm nt})$, where $P= B\Gamma$ and $B c$ is the speed of the plasmoid.  At
$x+\delta x$, the total x-component of the momentum of the plasmoid, including the swept-up particles and
radiated photons, is given by
$$\Pi_x(x+\delta x) = m_p(P+P^\prime\delta x)\{ (N_{\rm th}+ N_{\rm th}^\prime\delta x)(1+a_{\rm
th}+a_{\rm th}^\prime\delta x) +$$
$$+\int_0^\infty dp\; \gamma\; [N_p + N_p^\prime\delta x + (a_{\rm
nt}+a_{\rm nt}^\prime\delta x)(N_e+N_e^\prime\delta x)]\} -{\delta x \over c}\int_0^\infty
dp\; (\dot\gamma_p N_p +\dot\gamma_e N_e).\eqno(2)$$ 
The last integral in this equation refers to the momentum carried by radiated photons, and is related
to the internal particle distribution functions through the comoving frame energy-loss rates
$m_p\dot\gamma_p$ and $m_e\dot\gamma_e$ of the protons and electrons, respectively. This formalism allows
for particle creation and annihilation, although in this paper we assume for simplicity that $a_{\rm
th}^\prime = a_{\rm nt}^\prime = 0$.  We furthermore set $N_{\rm th}^\prime = 0$ so that no particles are
added to the thermal pool.  This does not impose any loss of generality if the nonthermal particles are
followed to nonrelativistic energies.

Expanding equation (2) in powers of $\delta x$, equating to $\Pi_x(x)$ by momentum conservation and
retaining terms to first order in
$\delta x$,  we obtain the equation
$$-{P^\prime\over P} = { \int_0^\infty dp\; \gamma (N_p^\prime + a_{\rm nt} N_e^\prime)-
(cP)^{-1}\int_0^\infty dp\; (\dot\gamma_p N_p +\dot\gamma_e N_e)  \over
N_{\rm th}(1+a_{\rm th}) + \int_0^\infty dp\; \gamma (N_p + a_{\rm nt} N_e)}\;.\eqno(3)$$

The proton and electron injection function due to the plasmoid sweeping up particles
from the ISM is given by $$N_{p,{\rm sw}}^\prime = n_{\rm ext}A\delta(p-P) = N_{e,{\rm
sw}}^\prime\;.\;\eqno(4)$$ 
Equation (4) says that protons and electrons are injected into the comoving
plasma with momenta $p = B\Gamma$ at a rate determined by the ISM density $n_{\rm ext}(x)$ and area $A(x)$
of the plasmoid.  When radiation is important, the functions $N^\prime_p$ and  $N^\prime_e$ in equations
(2) and (3) contain terms in addition to the sweep-up functions (4) due to the evolution of the comoving
particle distribution functions.  By multiplying the particle continuity equation by $\gamma$
and integrating over $d\gamma$, using the relation $dx = B\Gamma cdt$, we find that
$$\int_1^\infty  d\gamma \;\gamma\{\dot N(\gamma)+{\partial [\dot \gamma N(\gamma)]\over
\partial
\gamma}\} =
\int_1^\infty d\gamma \; [\gamma\dot N(\gamma)-\dot\gamma N(\gamma)] = B\Gamma c\int_1^\infty d\gamma\;\gamma
N_{\rm sw}^\prime\;, \eqno(5)$$ 
noting that $[\gamma\dot\gamma N(\gamma)]|_1^\infty = 0$. Consequently the numerator in equation (3) contains
only the contribution from the sweep-up function (4), and we arrive at the result
$$-{P^\prime\over P} = { \int_0^\infty dp\; \gamma (N_{p,{\rm sw}}^\prime + a_{\rm nt} N_{e,{\rm
sw}}^\prime) \over N_{\rm th}(1+a_{\rm th}) + \int_0^\infty dp\; \gamma (N_p + a_{\rm nt}
N_e)}\;.\eqno(6)$$
Note that we have neglected the energy content in magnetic field in equation (3) by assuming that it
is small in comparison with that of the thermal and nonthermal particle energies.  The importance of the
magnetic energy term can be checked once the plasmoid volume is specified, and can be important if
equipartition is assumed between the magnetic field and nonthermal particle energy densities. For
the applications presented here, however, we assume that this term is negligible. 

\section{Asymptotic Behavior of $\Gamma(x)$}

The relation between the comoving time $t$ and spatial coordinate $x$ is given by 
$${1\over c} \int_{x_0}^x \;{d\tilde x\over B(\tilde x)\Gamma(\tilde x)} = \int_{t_0}^t d\tilde t =
t-t_0\;.\;\eqno(7)$$
The terms $x_0$ and $t_0$ in equation (7) stand for the initial location and comoving time
when the parameter values are specified. The nonthermal power from swept-up protons and
leptons is probably dominated by the proton component in the ISM
environment surrounding a GRB source, though a substantial pair content could be found in the vicinity of
supermassive black holes which power blazars.  Assuming for simplicity that the swept-up matter is pair free, we
can therefore write that the power injected in nothermal particles per unit comoving time is given by
$\dot E = m_p\int_0^\infty dp\;\gamma\; \dot N_{p,{\rm sw}} =  m_p B\Gamma c\int_0^\infty dp\;\gamma\; 
N_{p,{\rm sw}}^\prime$.  Using equation (4), we therefore see that the kinetic energy injected
per unit time is 
$$\dot E_{\rm ke} =  m_p B(\Gamma^2-\Gamma) cn_{\rm ext}A\;\eqno(8)$$
(Blandford \& McKee \markcite{Blandford76}1976), neglecting the power in swept-up electrons.  Vietri
(\markcite{Vietri97a}1997a,\markcite{Vietri97b}b) has used this equation to predict GRB afterglow
behavior assuming that the bulk of the nonthermal particle energy is radiated on a time scale
short compared to the time scale for the evolution of $\Gamma$, i.e., that the plasmoid is in the
radiative regime.

When particle radiation is negligible, $\dot\gamma_p=\dot\gamma_e = 0$, and the evolution of the particle
distribution function is entire due to the
sweep-up nonthermal particles. Under these conditions, equation (6) reduces to
$$-{P^\prime\over P} \cong { n_{\rm ext}(x)A(x)\Gamma(x)\over
N_{\rm th} + \int_0^x d\tilde x\; n_{\rm ext}(\tilde x)A(\tilde x)\Gamma(\tilde
x)}\;.\eqno(9)$$
Equation (9) describes the dynamics of
the plasmoid in the {\it non-radiative} regime\footnote[1]{\rm This limit is also commonly referred to as
the adiabatic regime. We avoid use of this term which can be confused with the early
phases of the fireball expansion where adiabatic losses dominate the energy evolution of particles in the
comoving frame.}. If the particle kinetic energy is promptly radiated away, leaving behind only the
rest mass of the initial and swept-up particles -- which is dominated by the mass of the protons --
then equation (6) reduces to 
$$-{P^\prime\over P} \cong { n_{\rm ext}(x)A(x)\Gamma(x)\over
N_{\rm th} + \int_0^x d\tilde x \; n_{\rm ext}(\tilde x)A(\tilde x)}\;.\eqno(10)$$
This equation describes the dynamics of the plasmoid in the {\it radiative} regime.  

\subsection{Non-Radiative Regime}

Asymptotic forms for plasmoid dynamics in the non-radiative and radiative regime are simply
derived for a relativistic ($\Gamma \gg 1$) plasmoid.  We parameterize the plasmoid area by the
expression
$$A = A_0({x\over x_0})^j = 4\pi x_0^2 f_b ({x\over x_0})^j\;,\eqno(11)$$
where $f_b = \delta\Omega/4\pi$ is a collimation (or beaming) factor and $\delta\Omega$ is the
solid angle into which the plasmoid is directed. Equation (11) reduces to an uncollimated blast wave
when $j = 2$ and $f_b = 1$.  We parameterize the spatial dependence of the external density
distribution by the expression
$$n_{\rm ext}(x_i) = n_0({x\over x_0})^{-\eta}\;,\eqno(12)$$
and assume that the plasmoid Lorentz factor follows the power-law behavior
$$\Gamma(x) = \Gamma_0({x\over x_0})^{-g}\; ,\;\eqno(13)$$
where $\Gamma_0$ is the initial plasmoid Lorentz factor.

In the early free-expansion stage, the plasmoid is ballistic so that $\Gamma(x) = \Gamma_0$. 
Transition to power-law behavior in both the non-radiative and radiative regimes occurs when the relativistic mass
${\cal M}_{\rm sw}\Gamma_0$ of the swept-up particles equals the original mass $m_p N_{\rm th}$ in the burst
ejecta.  This occurs when 
$$\int_0^{x_0} d\tilde x\; n_{\rm ext}(\tilde x) A(\tilde x)\Gamma(\tilde x) \cong
\Gamma_0 \int_0^{x_0} d\tilde x\; n_{\rm ext}(\tilde x) A(\tilde x) \geq N_{\rm th} = {E_0\over
\Gamma_0 m_p}\;,\eqno(14)$$
where the equality on the right-hand-side relates the initial energy
$E_0 = 10^{50} E_{50}$ ergs of the GRB explosion or blazar flare to $N_{\rm th}$. One finds
that
$$x_0 = [{(j+1-\eta)E_0\over 4\pi f_b n_0\Gamma_0^2 m_p}]^{1/3} = 3.9\times 10^{15}
[{(j+1-\eta)E_{50}\over  f_b n_0\Gamma_{300}^2}]^{1/3}\;{\rm cm}\; ,\eqno(15)$$
where $\Gamma_{300} = \Gamma_0/300$. This reduces to the result of Rees \& M\'esz\'aros
(\markcite{Rees92}1992) when
$j = 2$, $\eta = 0$, and $f_b = 1$.  

Substituting equations (11)-(13) into equation (9), one obtains
$$-{\Gamma^\prime\over\Gamma} = {g\over x} = {j-g-\eta+1\over x}\;,\eqno(16)$$
so that equation (9) has a power-law solution in the non-radiative regime provided
$$g \rightarrow g_a = {j+1-\eta\over 2}\;.\eqno(17)$$
Note that the index $g_a$ in the non-radiative regime equals 3/2 for the blast-wave case with uniform external
density ($\eta = 0$). The power-law behavior persists until the plasmoid reaches location $x_f$
defined by
$\Gamma(x_f) \approx 1$.  For the asymptotic form (13), we have
$$x_f = \Gamma_0^{1/g} x_0\; . \eqno(18)$$

\subsection{Radiative Regime}

The plasmoid Lorentz factor in the fully radiative regime follows power-law behavior for
$x_0\lesssim x\lesssim x_f$, where $x_f$ is the location at which the mass of the swept-up matter
equals the initial rest mass of particles in the GRB ejecta. When $N_{\rm th} \gg \int_0^x d\tilde x\;
n_{\rm ext}(\tilde x) A(\tilde x)$, we can simply see from equation (10) that $-d\Gamma/\Gamma^2
\propto x^{j-\eta} dx$, so that 
$$g \rightarrow g_r = j+1-\eta = 2g_a\; .\eqno(19) $$
Alternately, we can solve equation (10) directly when the integral in the
denominator on the right-hand-side of equation (8) can be neglected. For a blast-wave geometry in a
uniform external medium,
$g_r = 3$ (Blandford \& McKee \markcite{Blandford76}1976; see also Rees \& M\'esz\'aros
\markcite{Rees92}1992; M\'esz\'aros et al.\ \markcite{MRW97}1997).  The power-law behavior in the
radiative regime ends at the location
$x_f$ determined by the condition
$$N_{\rm th} = \int_{x_0}^{x_f}d\tilde x\; n_{\rm ext}(\tilde x) A(\tilde x)\; .\eqno(20)$$
If $x_f \gg x_0$, this implies 
$${x_f\over x_0} = ({E_0 g_r\over \Gamma_0 m_p n_0 A_0 x_0})^{1/g_r}\; .\eqno(21)$$
The power-law behavior ends when $\Gamma(x_f) \approx 1$.  Using equations (18) and (19), we
therefore find that $x_0$ is also given by equation (15) in the radiative regime. Figure 2 shows
a numerical solution of $\Gamma(x)$ along with the asymptotic forms in the radiative and
non-radiative regimes for our standard parameter values
$\Gamma_{300} = 1$, $n_0 = 1$ cm$^{-3}$, $E_{50} = 1$, $j = 2$, $\eta = 0$, and $f_b = 1$ (see
Table 1).

\section{Energy Transfer from Protons to Electrons}

The bulk of the energy injected into a relativistic plasmoid which sweeps up matter from the ISM
is carried by nonthermal protons, yet in all likelihood the dominant radiation process in GRB
afterglows is nonthermal electron synchrotron radiation (see, e.g.,  Waxman \markcite{Waxman97a}1997a;
Tavani \markcite{Tavani97}1997; Katz \& Piran \markcite{Katz97}1997; and discussion in Chiang \& Dermer
\markcite{Chiang97a}1997a). It is also generally thought that the bulk of the radio-through-optical blazar
emission is nonthermal electron synchrotron radiation.  A mechanism transfers the energy from the proton
to the electron population, for example, through shocks or the excitation of plasma wave
turbulence by the protons which gets channeled into the electrons through gyroresonant coupling. 
We avoid the complicated microphysical details of the energy transfer process by offering two simple
prescriptions for the transfer of energy from the protons to the electrons.  

In the first prescription, we assume that a fraction $\xi_{pe}$ of the kinetic energy power
of the swept-up nonthermal protons in the comoving frame is instantaneously transformed into a
``shock"-like momentum power-law electron distribution with index $s$. Hence
$$\dot E_e = \xi_{pe}(x)\dot E = m_e K(x)\int_{p_{\rm min}}^{p_{\rm max}}dp\;(\gamma -1)
p^{-s}\;= m_e K(x) f .\eqno(22)$$
Naive shock acceleration theory predicts $s \geq 2$ and $s = 2$ for strong shocks in nonrelativistic
gases (e.g. Jones \& Ellison
\markcite{Jones91}1991).  Consideration of relativistic gases and inclusion of nonlinear effects can
produce harder power law injection with
$s > 1$ (e.g. Ellison, Jones, \& Reynolds \markcite{Ellison90}1990), and stochastic gyroresonant
acceleration can produce extremely hard power-law and even quasi-monoenergetic spectra (see
Schlickeiser, Campeanu, \& Lerche \markcite{Schlickeiser93}1993).  In equation (22), the normalization
coefficient
$K = \xi_{pe} \dot E/ f m_e$, where
$$f =  \int_{p_{\rm min}}^1 dp\;(\gamma -1) p^{-s} + \int_1^{p_{\rm max}}dp\;(\gamma -1) p^{-s}
\cong {1-p_{\rm min}^{3-s}\over 2(3-s)} + {p_{\rm max}^{2-s} - 1\over 2-s}\;,\eqno(23)$$
provided $p_{\rm min} \ll 1$ and $p_{\rm max} \gg 1$.  If $p_{\rm min} \gg 1$, then
$$f \cong  {p_{\rm max}^{2-s} - p_{\rm min}^{2-s}\over 2-s}\;.\eqno(24)$$

A second prescription for the energy transfer from protons to electrons is 
$$\dot E_e ={ E_{p,{\rm ke}}\over \tau_{pe}}\; , \eqno(25)$$
where 
$$E_{p,{\rm ke}} = m_p\int_0^\infty dp\;(\gamma - 1)N_p\; .\eqno(26)$$
Note that the energy transfer time scale $\tau_{pe}$ and the total proton kinetic energy $E_{p,{\rm
ke}}$ are $x$- (or equivalently $t$-) dependent quantities.  The second prescription is probably
a better representation of the actual physics of the situation but affords less analytic
development.  The limit of complete, prompt energy transfer from the protons to the
electrons corresponds to $\xi_{pe} \rightarrow 1$ and
$\tau_{pe}\rightarrow 0$ in the first and second prescriptions, respectively.  

The electron injection function at time $t_i$ in the comoving frame for the first prescription is
therefore
$$\dot N_e (p_i,t_i)= K\; p_i^{-s}
\Theta(p_i; p_{\rm min}, p_{\rm max})\; = \; {\xi_{pe} m_p c B (\Gamma^2 - \Gamma) n_{\rm ext} A
\over m_e f}\; p_i^{-s}
\Theta(p_i; p_{\rm min}, p_{\rm max})\; ,\eqno(27)$$
where $p_i$ is the injection momentum and $\Theta(p; p_{\rm min}, p_{\rm max})$ is the Heaviside
function defined by
$\Theta(x;x_1,x_2) = 1$ for $x_1\leq x < x_2$ and $\Theta(x;x_1,x_2) = 0$ otherwise.

The evolution of the electron energy through synchrotron radiation is governed by the well-known
equation
$$-\dot \gamma = {4\over 3}c\sigma_{\rm T} u_{\rm H}p^2\; = \nu p^2 \; , \;\;u_{\rm H} = {H^2\over
8\pi m_ec^2}\;,\eqno(28)$$
assuming an isotropic electron
pitch-angle distribution.  We also suppose that the magnetic field is randomly oriented
with mean field strength $H$ in our spectral calculations.  In general, $H = H(x)$.  The evolution
of $H$ with $x$ involves subtleties.  The assignment of an equipartition strength in GRB studies depends on
knowledge of the local density which can be obtained through the shock jump conditions (e.g.,
Waxman \markcite{Waxman97a}1997a; M\'esz\'aros et al.\ \markcite{MRW97}1997).  The magnetic field could be
generated through turbulent field growth, or  evolve from the original field in the plasmoid
(M\'esz\'aros et al.\  \markcite{MRP97}1994), for example, by flux freezing (see Chiang \& Dermer
\markcite{Chiang97a}1997a).  Our major simplification in the present study is to examine the case
$H = H_0$ and $\nu =\nu_0 = \sigma_T c H_0^2/(6\pi m_e c^2)$, with
$H_0$ and $\nu_0$ independent of $x$. We are also mostly interested in the radiation from relativistic
electrons ($p\gg 1$), in which case the electron equation of motion $-\dot p = \nu_0 p^2$ has the
solution
$$p = p(t) = [p_i^{-1} +\nu_0(t-t_0)]^{-1}\; \;\eqno(29)$$
(see Dermer \& Schlickeiser \markcite{Dermer93}1993 for a related treatment; Sturner, \&
Schlickeiser \markcite{Dermer97}1997 for a treatment of Coulomb, bremsstrahlung, and Compton losses in
addition to synchrotron losses; and Gould \markcite{Gould75}1975, Goldshmidt \& Rephaeli
\markcite{Goldshmidt94}1994, Dermer \& Skibo \markcite{DermerSkibo97}1997, and Chiang \& Dermer
\markcite{Chiang97a}1997a for a consideration of adiabatic energy losses of the nonthermal electrons). 
Here we treat only the effects of synchrotron losses on the evolution of the electron spectra.

The electron distribution at time $t$ resulting from the superposition of nonthermal electron
power-law injection functions described by equation (27) over the time interval $t_0 \leq t_i \leq
t$ is therefore
$$N_e(p;t) = \int _{t_0}^t dt_i \; \dot N_e(p,t_i) = p^{-2}\int_{{\rm
max}[t_0,t-\nu_0^{-1}(p^{-1}-p_{\rm max}^{-1})]}^t
dt_i\;K\;[p^{-1}-\nu_0(t-t_i)]^{s-2}\;,\eqno(30)$$
where we assume that $p_{\rm min}$ and $p_{\rm max}$ are independent of $t$.  We note that a value for
$p_{\rm max}$ can be obtained in shock acceleration theory by balancing radiative loss and energy gain
rates (e.g., Reynolds \markcite{Reynolds96}1996).

We now follow the evolution of the nonthermal electron distribution that occurs over the period
when the plasmoid Lorentz factor is described by the power-law asymptote (13).  Integrating
equation (7) from the onset of the power-law behavior at $x_0$ to location $x_i$, we find that
$${x_i\over x_0} = [\omega(t_i-t_0)+1]^{1/(1+g)}\;,\eqno(31)$$
where $\omega = (g+1)\Gamma_0 c/ x_0$. In the range $x_0 \leq x_i \leq x_f$, $\Gamma$, $A$, and
$n_{\rm ext}$ are described by equations (13), (11), and (12), respectively.  Thus
$$K \cong {\xi_{pe} m_p c \Gamma_0^2 n_0 A_0 \over m_e f} ({x_i\over
x_0})^{j-2g-\eta}\;,\eqno(30)$$
when $\Gamma \gg 1$.  

In this regime, equation (27) becomes
$$N_e(p;\tau ) = {\xi_{pe} m_p c \Gamma_0^2 n_0 A_0 \over m_e f p^2}\;
I_s(\tau)\;\eqno(33)$$
where
$$I_s(\tau) = \int_0^{{\rm min}[\tau,\nu_0^{-1}(p^{-1}-p_{\rm max}^{-1})]}
dy\;(1+\omega\tau-\omega y)^u\;(p^{-1}-\nu_0 y)^{s-2}\;.\eqno(34)$$
Here $\tau = t-t_0$ and $u = (j-2g-\eta)/(1+g)$ (the integration
variable $y = t-t_i$).  The simplest case is given by $s = 2$, which corresponds to the strong
shock index in the quasilinear regime of shock acceleration theory. When $s=2$, equation (34)
becomes
$$I_2(\tau) =   \cases{  [\omega(1+u)]^{-1}[(1+\omega\tau)^{c_0} - (1+\omega\tau-\omega\bar
y)^{c_0}],& if
$c_0 \neq 0$;\cr\cr 
\omega^{-1} \ln({1+\omega\tau\over 1+\omega\tau - \omega \bar y}),& if $c_0 = 0$,\cr}\eqno(35)$$
where $\bar y = {\rm min}[\tau,\nu_0^{-1}(p^{-1}-p_{\rm max}^{-1})]$ and $c_0 = 1+
u = (g_r-g)/(g+1)$, recalling definition (19).

\section{Synchrotron Radiation}

The synchrotron emissivity in the comoving frame is treated in the $\delta$-function approximation
of Dermer \& Schlickeiser (\markcite{Dermer93}1993; eq.[5.16]), which we write as
$$\dot N_{\rm syn,com}(\e,\tau) = {\nu_0\over \e} \;\int_0^\infty
dp\;p^2\;N_e(p;\tau)\delta[p-({\e\over\eh})^{1/2}] = {\nu_0\over
2\eh^2}({\eh\over\e})^{1/2}\;N_e[({\e\over\eh})^{1/2};\tau]\;.\eqno(36)$$
The dimensionless cyclotron energy $\eh = H/H_{\rm cr}$, where $H_{\rm cr} = 4.414\times
10^{13}$ Gauss.  For a power-law electron distribution, approximation (36) is accurate
everywhere except near the endpoints of the synchrotron spectrum.  Self-absorption of the
synchrotron radiation and synchrotron self-Compton emission (see Dermer, Sturner, \&
Schlickeiser \markcite{Dermer97}1997 for a treatment of the latter process) are not considered here.

The relationship between the observer's time element $\delta \to$ and the comoving time element
$\delta t$ is $\delta \to = \delta t (1+z)/\D$, where the Doppler factor $\D =
[\Gamma(1-B\mu_{\rm obs})]^{-1}$ and $\arccos \mu_{\rm obs}$ is the angle between the direction of
travel of an emission element and the observer's line-of-sight.  We consider the case where
a portion of the plasmoid is directed along the line-of-sight to the observer, so that the
received radiation is produced primarily by emitting plasma which is directed within a line-of-sight
angle
$\arccos \mu_{\rm obs}\lesssim 1/\Gamma$ (see Chiang \& Dermer
\markcite{Chiang97a}1997a when this is not the case).  Thus we can let $\delta \to \rightarrow \delta
t(1+z)/\Gamma$.\footnote[2]{\rm At a fixed observer's time, 
the mean value of $\mu$ for a shell which emits
uniformly per unit surface and expands with constant Lorentz factor is 
equal to $B$ and $4B/(3+B^2)$ for
photons emitted isotropically in the comoving frame 
with energy index $\alpha$ equal to 0 and 1, respectively. 
When $\mu_{\rm obs} \approx B$, $\D \cong \Gamma$.}

The observed photon energy (in units of $m_e$) is given by  $\eo  = \D\epsilon/(1+z)\rightarrow
\Gamma\epsilon/(1+z)$, where the last relation again holds for the emitting plasmoid directed
within an angle $\arccos \mu_{\rm obs}\lesssim 1/\Gamma$ of the observer's line-of-sight.  The
number of photons produced per unit observer's time per unit observed photon energy $\eo$ is
therefore given by
$$\dot N_{\rm syn}(\eo,\to) = \dot N_{\rm syn,com}[\e(\eo),t(\to)]\;,\eqno(37)$$
noting that the Jacobian is unity.  The quantity $\dot N_{\rm com}(\e,t)$ is the
differential photon production rate in the comoving frame. 

In equation (37), we replace $\e$ by $(1+z)\eo/\Gamma$ and $t$ by a function of $\to $
defined through the expression
$$\to -t_{\rm obs,0} = \tauo = (1+z)\int_{t_0}^t {d\tilde t\over \Gamma(\tilde t)}\;.\eqno(38)$$
In the power-law regime for the plasmoid Lorentz factor, equations (38), (13) and (31) imply
$$\omega\tau = (\Omega\tauo + 1)^{c_1} - 1\;,\eqno(39)$$
where $\Omega = \omega\Gamma_0(1+2g)/[(1+z)(1+g)] = (2g+1)c\Gamma_0^2/[x_0(1+z)]$ and $c_1 =
(1+g)/(1+2g)$.

\section{Numerical Results}

We may now calculate the observed synchrotron radiation spectrum using equations (33) and (36) in
equation (37).  The power-law solution for $\Gamma$ becomes accurate after observer's time
$$\to \gtrsim \Omega^{-1} \cong 1.4\; {1+z \over \Gamma_{300}^{8/3} (2g+1) }\;({g_r
E_{50}\over f_b n_0})^{1/3} \;\;{\rm sec}\;\;\eqno(40)$$  
following the start of the GRB, which
represents only a couple seconds for our standard parameters (see eq.[15] and Table 1).  A value of
$p_{\rm max} = 10^{7}/H_0^{1/2}$ is used, which is consistent with a maximum synchrotron photon energy
$\lesssim 25$ MeV in the comoving frame of
 expected from diffusive shock acceleration (de Jager \& Baring
\markcite{deJager97}1997). The solution is no longer valid when $\Gamma
\approx 1$, and we stop our calculations there. 

Figure 3 illustrates the relationship between $\Gamma/\Gamma_0$ and $x/x_0$ as a function of the quantity
$\Omega \to$. As expected, a highly radiative blast wave in a uniform external medium (i.e., $g=3$) 
slows down most rapidly and travels the shortest distance.  When $g < 3$, due either to a non-blast
wave geometry with $j < 2$, a thinning external medium with $\eta > 0$, or  inefficient radiation of the
swept-up nonthermal proton energy, the plasmoid decelerates more slowly.

The spectra at different observing times and the time profiles at different observed photon
energies are presented in Figures 4-6.  The spectra are plotted at observing times $\to = 1$, 10,
$10^2$ sec, etc., from top to bottom.  The spectra show a spectral hardening from photon energy index
$\alpha = 1/2$ at lower photon energies to $\alpha \gtrsim 1$ at higher photon energies due to the effects
of cooling on the spectra for the index $s = 2$ used in the electron momentum spectrum.  Spectra
measured at late times display a spectral break at lower photon energies because the
effects of cooling on the lower energy electrons are only felt at late times.  A pile-up appears
at the spectral break, and this pile-up becomes more pronounced at later times and lower photon energies
as more electrons are swept into the pile-up region. 

Figure 4 shows the spectra and time profiles using all standard parameters (see Table 1).  The
standard value of 10 Gauss for the magnetic field is arbitrarily assigned in order to provide a
spectral break at $\approx 100$ keV - several MeV, as is typical for many GRBs (e.g., Band et al.
\markcite{Band93}1993). The parameters in Figure 5 are the same as those in Figure 4 except that $H_0 = 1$
Gauss.  This has the effect of moving the energy of the spectral break to higher energies at comparable
observing times. Figure 6 shows a calculation using all standard parameters except for $\Gamma_0 =30$,
$n_0 = 0.01$ cm$^{-3}$, and a beaming factor $f_b = 0.01$.  This choice delays the onset of the
asymptotic power-law regime to late times (compare eq.[40]), and is driven by an attempt to explain
observations of the peaking of the optical and X-ray afterglow emission of GRB 970508 at $\approx 2$ days
following the burst.

The time profiles of Figures 4-6 can be understood qualitatively as follows:  The highest photon
energies are produced by very energetic electrons. In the comoving frame, only the most recently
injected high-energy electrons have not been degraded to lower energies by
synchrotron cooling.   The time profile of the decay at MeV - GeV energies is therefore determined
by the power in recently injected electrons, which reflects the time-dependence of the
power in the swept-up matter as the plasmoid decelerates.  The time profiles in this
regime follow a power law behavior $\propto \to ^{-\chi}$, with slope
$\chi\approx 10/7$, as explained in the next section.  The electrons which
produce synchrotron radiation at lower observed photon energies take much longer to cool.  Before
cooling is important, these electrons accumulate.  The time profiles in this regime consequently
decay more slowly than in the strongly cooled regime. The slope of the time profile reflects both the
accumulation of newly injected electrons and Doppler effects on the observer.  Because the
plasmoid is decelerating, synchrotron emission at a fixed observing energy is produced
increasingly by the fewer higher energy electrons.

The spectra in Figs. 4-6 are presented in the dimensionless form $\nu L_\nu/m_e$, and reach values of
$10^{52}-10^{54}$ during the first 1-100 seconds of the GRB.  The weakest GRBs observed with BATSE are
triggered at thresholds of $\approx 0.2$-1 photon cm$^{-2}$ s$^{-1}$ in the 50-300 keV range.  For a mean
photon energy of 100 keV for GRB sources at a distance of $10^{28}$ cm, roughly corresponding to $z =
1$, we would therefore expect $\nu L_\nu/m_e$ fluxes of $\approx (0.2-1)\times 4\pi\times 10^{56}\times
0.1/0.511
\cong 5\times 10^{55} - 3\times 10^{56}$. Consequently, the energetics can be explained only if a typical
GRB involves at least $10^{52}$ ergs or if $f_b \lesssim 0.01$ and $\Omega^{-1} \sim 1$- 10 sec.

There is a characteristic time $\hat\tau_{\rm obs}$ when synchrotron cooling starts to influence the total
number of electrons which produce radiation at a given energy. The value of $\hat\tau_{\rm obs}$  is
longer at lower photon energies, and this effect causes the X-ray to
$\gamma$-ray flux ratio to rapidly increase during the first 100 seconds in Figures 4 and
5.  For this model with constant magnetic field, the light curve of the decaying gamma-ray
emission asymptotically approaches a power-law with slope characteristic of the cooled
regime.  
 
\section{Analytic Interpretation}

Equation (39) can be rewritten in terms of a dimensionless time $\cross = 1+\Omega\tauo$ through  the
expression
$$1+\omega \tau = \cross^{c_1}\, .\eqno(41)$$
From equations (13) and (31),
$$\Gamma = \Gamma_0\cross^{-c_2}\, , \eqno(42)$$
where $c_2 = g/(2g+1)$.  We also write
$$p^2 = {\e\over\eh} = [{(1+z)\eo\over \Gamma_0\eh}]\;\cross^{c_2} = {\E}\cross^{c_2}\; .\eqno(43)$$ 
Equation (37) becomes, with equation (33) and (36),
$$\dot N_{\rm syn}({\E},\cross ) = {\nu_0 K_0\over
2\eh^2\Omega}\;{\E}^{-3/2}\cross^{-3c_2/2}I_s(\cross )\; ,\eqno(44)$$
where $$K_0 = {\xi_{\rm pe} m_p \eh c\Gamma_0^3 n_0 A_0\over (1+z) m_e f}\; .\eqno(45)$$

In the uncooled regime, $\tau \ll \nu_0^{-1}(p^{-1}-p_{\rm max}^{-1})$.  Examination of equation
(34) shows that when $p\ll p_{\rm max}$,
$$I_s(\cross ) \rightarrow I_s^{\rm uncooled} = {p^{2-s}\over c_0 \omega }  (\cross^{c_0\cdot c_1 }
-1) \; .\eqno(46)$$ 
Equation (45) applies for all values of $c_0\neq 0$; when $c_0 = 0$, $I_s^{\rm
uncooled} =
\omega^{-1}c_1\ln \cross$.  The time-dependence of the energy spectrum of the
synchrotron radiation in the regime where the emission is produced mainly by electrons which
have not cooled through synchrotron radiation is given by 
$$\dot N_{\rm syn}^{\rm uncooled}(\E ,\cross ) = {\nu_0 K_0\over
2c_0\eh^2\Omega\omega}\;\E^{-(s+1)/2}\; \cross^{-c_2(s+1)/2}\; (\cross^{c_0\cdot c_1}-1 )\;
.\eqno(47)$$

The regime where the bulk of the injected electrons has already cooled is characterized by the
condition $\tau \gg \nu_0^{-1}(p^{-1}-p_{\rm max}^{-1})$.  When $p\ll p_{\rm max}$, 
$$I_s(\cross ) \rightarrow I_s^{\rm cooled} = {p^{1-s}\over (s-1) \nu_0 }  \; \cross^{c_1\cdot (c_0 - 1)
} \; ,\eqno(48)$$
and the observed synchrotron flux is given by
$$\dot N_{\rm syn}^{\rm cooled}(\E ,\cross ) = {K_0\over
2(s-1)\eh^2\Omega}\;\E^{-(s+2)/2}\; \cross^{-c_2(s+2)/2\; + \; c_1\cdot (c_0 -1)} \; .\eqno(49)$$

From the above results, we see that the negative of the slopes of the synchrotron flux time
profiles produced by a decelerating plasmoid in the asymptotic power law regimes $\to \gg
\Omega^{-1}$ are given by
$$\chi_{\rm uncooled}  = -c_0\cdot c_1 + c_2(s+1)/2 \rightarrow c_2(s+1)/2\; ,\eqno(50)$$
and
$$\chi_{\rm cooled}  = -c_0\cdot c_1 + c_1 + c_2(s+2)/2 \rightarrow c_1 + c_2(s+2)/2\; ,\eqno(51)$$
in the uncooled and cooled regimes, respectively.  The expressions on the right-hand-side of
equations (50) and (51) represent the fully radiative regime where $g = g_r$ and, therefore, $c_0
= 0$.  Table 2 shows the derived slopes of the flux time profiles for emission in the fully
radiative uncooled and cooled regimes
When $g < g_r$, which happens when the incoming energy is not fully radiated -- and which must be
the case in view of the fact that there are still some nonthermal electrons which are radiating
energy -- the slopes are less negative than given here, so that the profiles decay more slowly.  A
numerical calculation (Chiang \& Dermer \markcite{Chiang97b}1997b) is required to determine the slopes of
the time profiles in this regime.  In the extreme non-radiative regime, where $g = g_a = g_r/2$, the amount of
radiated energy is negligible, and the slopes of the time profiles are harder by 
$c_0\cdot c_1 = g/(2g+1)$ units.  The derived slopes in the fully non-radiative regime are also listed
in Table 2, as well as the difference in slopes between the cooled and uncooled regimes. These differences
are the same in the radiative and non-radiative limits.

The transition from the regime where synchrotron cooling plays a negligible role to the regime
where it plays a major role in determining the number of nonthermal electrons with momentum $p$
occurs at comoving time $\hat\tau \approx (\nu_0 p)^{-1}$, provided $p \ll p_{\rm max}$.  This also
yields the observer time $\hat \tauo$ when synchrotron cooling starts to strongly regulate the
number of electrons which produce synchrotron emission at some observed energy.  For times
$\hat\tau
\gg \omega^{-1}$ or, equivalently, $\hat\tauo \gg \Omega^{-1}$, we can use equation (41) to write
$\hat\tau =\omega^{-1}(\Omega\hat\tauo)^{c_1}$ and equation (43) to write $p$ in terms of the
observed photon energy $\eo$.  After a tedious derivation, one obtains the result
$$\hat\tauo = { (1+z)^{1-c_3} (g+1)^{2c_3} 433^{1-2c_3} (4.1\times 10^6)^{c_3}\over 300
(2g+1) H_0^{3c_3}\eo^{c_3}}\;\Gamma_{300}^{(13c_3 -8)/8}\; ({g_r E_{50}\over f_b
n_0})^{(1-2c_3)/3}\;.\eqno(52)$$
The term $g_3 = 1/(2c_1 + c_2) = (2g+1)/(3g+2)$.  When $g = 3$,
$$\hat\tauo \cong 8.5\;{(1+z)^{0.364}\Gamma_{300}^{0.034}\over H_0^{1.91} \eo^{0.636} }\; ({f_b n_0 \over
E_{50} g_r})^{0.0909}\;\;{\rm sec}\;.\eqno(53)$$
When $g = 1$, 
$$\hat\tauo \cong 7.0\;{(1+z)^{0.4}\over H_0^{1.8} \eo^{0.6} \Gamma_{300}^{0.025} }\; ({f_b n_0 \over
E_{50} g_r})^{0.0667}\;\;{\rm sec}\;.\eqno(54)$$
The time at which synchrotron cooling effects begin to play an important role are very weakly dependent
upon all parameters except $H_0$ and the observed photon energy $\eo$. The accuracy of equation (53)
can be checked against Figs. 4 and 5 using $g_r = g = 3$.  In Fig. 6, the M\'esz\'aros-Rees timescale
$\Omega^{-1}$ is longer than $\hat\tauo$, so that synchrotron cooling effects are always important by the
time the plasmoid Lorentz factor enters the power-law asymptote regime described by equation (13).

\section{Discussion}

\subsection{Application to Gamma Ray Bursts}

We use our results to make preliminary comparison with observations of delayed afterglow behaviors of GRBs
noting, however, that a detailed numerical scheme is required to calculate accurately the time profiles of
the afterglows.  It seems likely that the X-ray and
optical GRB afterglow emission is primarily observed in the regime where synchrotron cooling plays an
important role in the nonthermal electron content which produces this radiation (see eqs.[53]
and [54]).  From Table 2, we see that the slopes of the time profiles in this ``cooled" regime, whether
in the non-radiative limit or the radiative limit, are fairly narrowly bracketed.  Irrespective of the
electron injection index
$s$ in the range $1 < s < 3$, the slopes $\chi$ lie between the values of $11/14 = 0.78$ and $23/14 =
1.64$.  If the optical and X-ray afterglow indeed occur in the regime where synchrotron cooling is
important, then $s = 2\alpha$ (see eq.[49]).  For GRB 970228, $\alpha \cong 0.63$ (e.g. Waxman
\markcite{Waxman97a}1997a), and for GRB 970508, $\alpha \cong 0.65$ (Djorgovski et al.\
\markcite{Djorgovski97}1997), implying $s\cong 1.3$ in both cases.  Such a spectrum can be produced
by nonlinear effects in a relativistic shock (Ellison et al. \markcite{Ellison90}1990). The radiative
regime with $g = g_r$ is suggested on energetic grounds (Vietri \markcite{Vietri97b}1997b).  From Table
2, we see that a relativistic blast wave decelerating in a uniform medium yields temporal slopes in the
range $17/14 = 1.21 \lesssim \chi \lesssim 37/28 = 1.32$ when $1 < s < 1.5$. These values are in striking
accordance with the slopes of the measured afterglow behaviors (see \S 1), 
particularly in view of
an additional hardening of the temporal slopes expected from the fact that the radiative regime can never
fully be achieved.

Against the apparent successes of the relativistic blast wave model in explaining the slopes of the
afterglow time profiles, several difficulties must be mentioned: (1) The
characteristic GRB spectral shape cannot be reproduced with the model presented here, which gives a 1/2
power break due to incomplete synchrotron cooling rather than a more typical break of about one unit from
the hard X-rays to the soft gamma rays.  This problem can be ameliorated by the inclusion of relativistic
shock effects which produce time-varying equipartition magnetic fields and shock-heated low-energy cutoffs
in the electron momentum distribution, and will be dealt with elsewhere (Chiang \& Dermer
\markcite{Chiang97b}1997b). (2) The short time-scale variability and the large diversity of GRB
time profiles are not accounted for here, and probably cannot result from variations in the external medium
(Fenimore, Madras, \& Nayakshin \markcite{Fenimore96}1996; Sari \& Piran 
\markcite{Sari}1995), though internal shocks may suffice (see Kobayashi, Piran, \& Sari
\markcite{Kobayashi97}1997). (3) The tendency of the peak in the
$\nu F_\nu$ spectrum of GRBs to be in the range 0.1-1 MeV -- which is probably not due to selection effects
(Harris
\& Share \markcite{Harris98}1998) -- remains unexplained by this model. (4) Finally, the peaking of the
X-ray and optical time profiles two days after GRB 970508 is unexpected.  Fig. 6 shows extreme
parameters which produce a delayed peaking of the X-ray and optical ranges, but implies an accompanying
long term gamma-ray production phase.  The X-ray and optical feature at 2 days after GRB 970508 must
therefore originate from another effect, for example by relativistic beaming due to a collimated plasmoid
(Chiang \& Dermer 1997a) or by a density enhancement which the expanding shell traverses as it decelerates.

\subsection{Application to Blazars}

The most extreme flaring behavior of blazars occurs in the gamma-ray regime, with luminosities $L_f$ 
exceeding $10^{48}L_{48}f_b$ ergs s$^{-1}$  and gamma-ray variability taking place on
time scales $t_f \lesssim t_{\rm day}$ days, with $L_{48} \sim 1$ and $t_{\rm day}\lesssim 1$. 
The Eddington limit for a $10^8 M_8 M_\odot$ black hole is
$\approx 10^{46}M_8$ ergs s$^{-1}$.  If the jet luminosity represents a significant fraction of the
Eddington luminosity and is collimated into 0.1 - 1\% of the full sky, then $10^8 M_\odot$ black holes
could power such flares.  This interpretation is also consistent with a  minimum variability time scale
defined by the Schwarzschild radius of the black hole, which is about 15 minutes when $M_8 \sim 1$.

Interactions of the jet plasma with clouds in the external medium could provide a mechanism for
transforming the kinetic energy of the jet power into the emitted radiation. Certain conditions have to
be met for this interpretation to be valid. In particular, the clouds have to be (i) small enough to
be in accord with the observed variability time scale, (ii) dense and massive enough to afford efficient
conversion of the plasmoid energy into radiation, (iii) close enough to the central black hole so that a
cloud can pass through the volume traversed by the jet on a time scale of order days, (iv) large enough to
occult a large portion of the jet, and (v) numerous enough to agree with the blazar flare duty cycle.  

Neglecting redshift factors in these estimates, we see that condition (i) implies that $r_{\rm cloud}/
c\Gamma^2 \lesssim t_f$, so that the cloud radius
$r_{\rm cloud} \lesssim 3\times 10^{17} \Gamma_{10}^2 t_{\rm day}$ cm, where $\Gamma_{10} = \Gamma/10$.
Assuming that the plasmoid is highly radiative, equation (8) gives $4\pi f_b  d_{\rm cloud}^2
m_p\Gamma^2 c n_{\rm cloud} \gtrsim 10^{48}L_{48}f_b$,  where $d_{\rm cloud} = 10^{16}d_{16}$ cm is
the distance of the cloud from the supermassive black hole.  Condition (ii) therefore implies that
$n_{\rm cloud}$ $\gtrsim 2\times 10^5 L_{48}(\Gamma_{10}d_{16})^{-2}$ cm$^{-3}$.   In the fully
radiative limit, we see from equation (20) that a plasmoid will be decelerated to nonrelativistic
speeds if the number of swept-up particles roughly equals the number of jet particles.  Consequently
we find that the cloud mass
$M_{\rm cloud} \cong M_{\rm plasmoid} \cong L_f t_f/\Gamma$, so that $M_{\rm cloud} \gtrsim 10^{52}
L_{48}f_b t_{\rm day}/\Gamma_{10}$ ergs. For a spherical cloud, we therefore see that $n_{\rm cloud}
\gtrsim 10^9 L_{48}f_b t_{\rm day}/R_{15}^3\Gamma_{10}$ cm$^{-3}$, where we define $R_{15} = r_{\rm
cloud}/(10^{15}{\rm cm})$.

For condition (iii), we find that the cloud takes  $\approx 1.7
d_{16}^{3/2}/M_8^{1/2}$ days to travel one degree if the clouds follow circular orbits, and could
therefore traverse the jet opening angle on a time scale of several days for the given parameters.
Condition (iv) is satisfied if
$R_{15}\gtrsim d_{16}(\theta_j/0.1{\rm~ rad}$), where
$\theta_j$ is the jet opening angle. The duty cycle of flares is not well known due to the sparse sampling
of blazars, but might be in the range of 1-10\%.  Condition (v) therefore translates into a covering factor
of comparable value as the duty cycle.

The interpretation that blazar flares could result from relativistic plasma outflows interacting
with clouds is therefore consistent with observations if marginally Thomson thick clouds of radii $\sim
10^{15}$ cm are located some hundreds to thousands of Schwarzschild radii from the central supermassive
black holes.

\section{Summary}

A method is presented in this paper to treat the dynamics of a volume of relativistically moving
magnetized plasma -- a plasmoid -- that sweeps up material from the interstellar medium and whose
entrained nonthermal particles radiate their internal kinetic energy, thereby changing the relativistic
inertia of the plasmoid.  The formalism is cast in terms of particle distribution
functions in the comoving plasmoid frame: these functions evolve when particles are swept-up, when energy
is transferred between the hadronic and leptonic components in the plasmoid, and when energy is lost to
radiation.  Such an approach is open-ended in the sense that more complicated scenarios not dealt with here
can be treated, such as the inclusion of other radiation processes (e.g., Compton scattering
and bremsstrahlung, and secondary and photo-meson production), of diffusive
escape of nonthermal particles from the plasmoid, and of pair and compactness effects and the microphysical
details of particle acceleration within the plasmoid. 

We have studied a highly idealized system consisting of a relativistic blast wave or plasmoid with a
constant entrained magnetic field.  Only optically-thin synchrotron processes were treated here, and
it was assumed that a fixed fraction of the swept-up nonthermal proton energy is instantaneously
transformed into a power-law nonthermal electron momentum spectrum.  This model was completely solved
in the limit of a relativistic plasmoid whose Lorentz factor follows the power-law behavior given by
equation (13). This solution can be used to benchmark numerical codes that include additional
processes.  Application of these results to studies of gamma-ray bursts and blazars was indicated, and
will be developed more thoroughly in a numerical examination of the system.

\acknowledgements

We thank M. B\"ottcher for criticisms which helped clarified the derivation of equation (6), and
we acknowledge useful comments by M. G. Baring, A. K. Harding, and F. C. Jones.  The work of CDD
was supported by the Office of Naval Research.  The work of JC was performed while he held a
National Research Council - Naval Research Laboratory Associateship.

\begin{table}[h]
\small
\caption{List of Parameters, their Standard Values,
and Values of Derived Quantities}
\bigskip
\begin{tabular}{c l c}
\tableline\tableline
Parameter or & Standard Value \\
Derived Quantity &  \\
\tableline
$\Gamma = \Gamma_0 (x/x_0)^{-g}$ & g = 3 \\
$\Gamma_0 = 300 \Gamma_{300}$ & $\Gamma_{300} = 1$ \\
$n_{\rm ext} = n_0 (x/x_0)^{-\eta}$  & $\eta  = 0$ \\
$n_0$  & $ 1$ cm$^{-3}$\\
$E_{50} = E_0/10^{50}$ ergs & $E_{50} = 1$ \\
$A = A_0 (x/x_0)^{j}$ & $j = 2$ \\
$f_b$ & $ 1$ \\
$H_0$ & 10 Gauss \\
$s$ & 2 \\
$z$ & 1 \\
 & \\
$g_r = j+ 1 - \eta$ & 3 \\
$g_a = (j+1 -\eta)/2$ & 3/2 \\
$x_0$ & 0.0018 pc \\
$A_0 = 4\pi x_0^2 f_b$ & $3.9\times 10^{32}$ cm$^{-2}$\\
$\omega^{-1}$ & 156 sec \\
$\Omega^{-1}$ & 0.58 sec \\
\tableline
\end{tabular}
\label{table1}
\end{table}

\begin{table}[h]
\small\centering
\caption{Negative of Slopes of Time Profiles in Fully Radiative Regime $g = g_r = j+1 -
\eta $ and Fully Non-Radiative Regime $g = g_a = g_r/2$ }
\bigskip
\begin{tabular}{ccccccccccccc}
\tableline\tableline
 &   & s = 1 & &  & $s = 1.5 $ & &  &$s = 2$ & &  & $s = 3$  & \\
 g~~~~ & $ 1$ & 2 & 3~~~~ & $ 1$ & 2 & 3~~~~ & 1 & 2 & 3 ~~~~& 1 & 2 & 3 \\
\tableline
$c_1$~~~~ & $2/3$ & $3/5$ & $4/7$~~~~ & $2/3$ & $3/5$ & $4/7$~~~~ & $2/3$ & $3/5$ & $4/7$~~~~ &
$2/3$ & $3/5$ & $4/7$ \\
$c_2$ ~~~~& $1/3$ & $2/5$ & $3/7$ ~~~~& $1/3$ & $2/5$ & $3/7$~~~~ & $1/3$ & $2/5$ & $3/7$ ~~~~&
$1/3$ & $2/5$ & $3/7$ \\
$c^{\rm rad}_{\rm uncooled}$ ~~~~& $1/3$ & $2/5$ & $3/7$~~~~& $5/12$ & $1/2$ & $15/28$~~~~ & $1/2$ &
$3/5$ & $9/14$ ~~~~&
$2/3$ & $4/5$ & $6/7$ \\
$c^{\rm rad}_{\rm cooled}$ ~~~~& $7/6$ & $6/5$ & $17/14$~~~~& $5/4$ & $13/10$ & $37/28$~~~~ & $4/3$ &
$7/5$ & $10/7$ ~~~~&
$3/2$ & $8/5$ & $23/14$ \\
$c^{\rm adi}_{\rm uncooled}$ ~~~~& $0$ & $0$ & $0$~~~~& $1/12$ & $1/10$ & $3/28$~~~~ & $1/6$ &
$1/5$ & $3/14$ ~~~~&
$1/3$ & $2/5$ & $3/7$ \\
$c^{\rm adi}_{\rm cooled}$ ~~~~& $5/6$ & $4/5$ & $11/14$~~~~& $11/12$ & $9/10$ & $25/28$~~~~ & $1$ &
$1$ & $1$ ~~~~&
$7/6$ & $6/5$ & $17/14$ \\
$\Delta c$ ~~~~& $5/6$ & $4/5$ & $11/14$~~~~& $5/6$ & $4/5$ & $11/14$~~~~ & $5/6$ & $4/5$ & $11/14$ ~~~~&
$5/6$ & $4/5$ &
$11/14$ \\
\tableline
\end{tabular}
\label{table2}
\end{table}

\eject

\begin{figure}
\centerline{\epsfysize=10cm \epsfbox{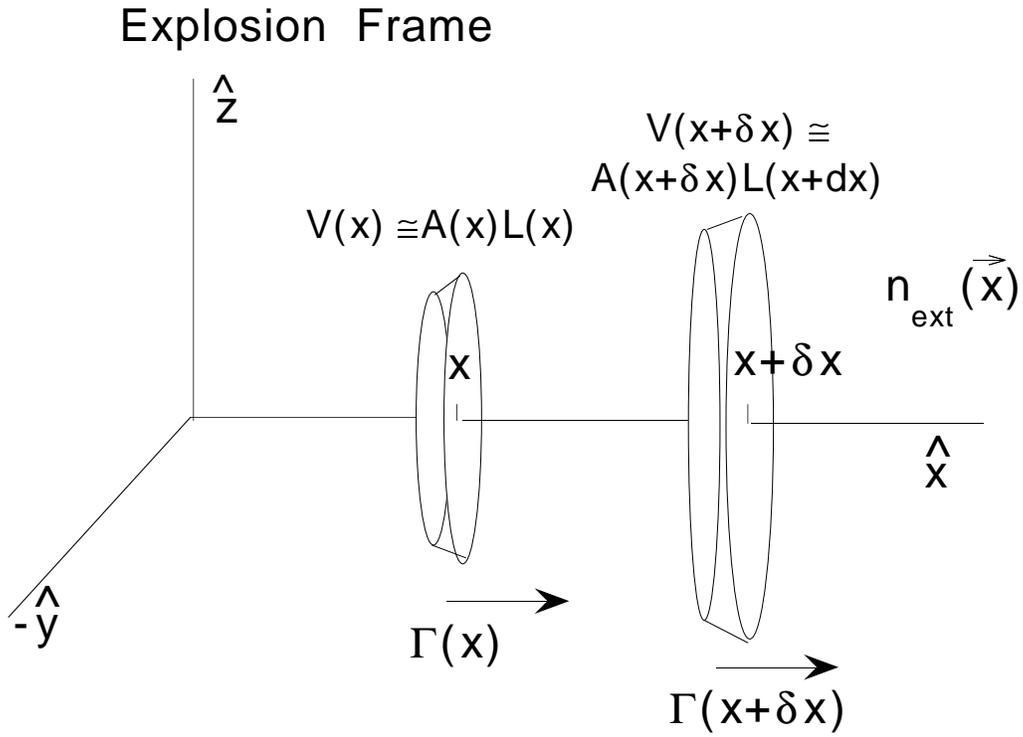}}
\caption{Cartoon illustrating the geometry of the plasmoid outflow and deceleration. 
The variation of the plasmoid bulk Lorentz factor $\Gamma(x)$ with location $x$ depends on the
amount of swept-up external matter, which is characterized by density $n_{\rm ext}({\bf x})$, and the
amount of internal energy which is radiated.}
\end{figure}

\begin{figure}
\centerline{\epsfysize=25cm \epsfbox{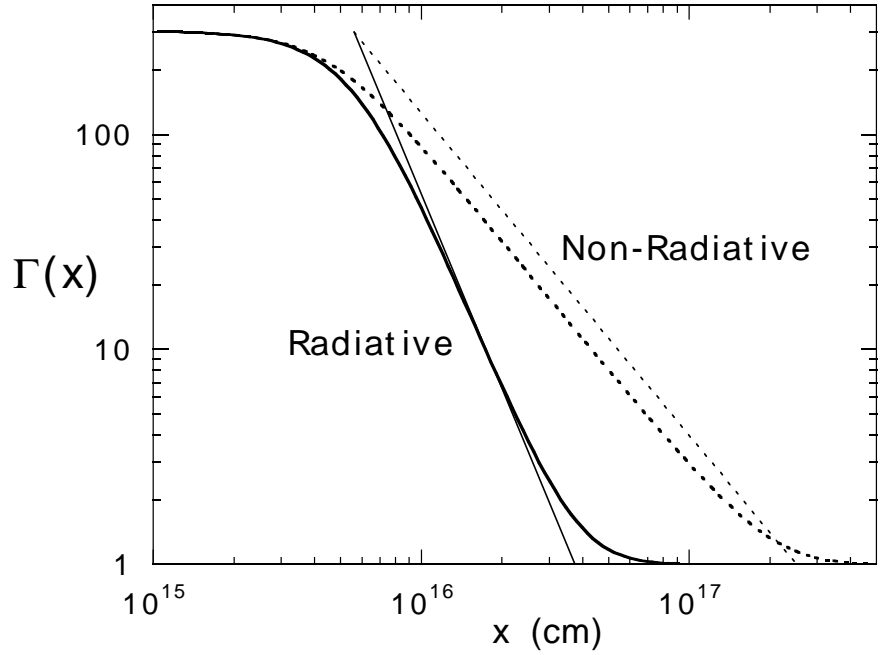}}
\vspace{-3.0in}
\caption{Solid and dotted curves show numerical solutions of equation (6) for the
plasmoid Lorentz factor $\Gamma$ as a function of location $x$ in the non-radiative and radiative
regimes, respectively.  Standard parameter values, shown in Table 1, are used. Solid and
dotted lines give the power-law asymptotes in the respective regimes.}
\end{figure}

\begin{figure}
\centerline{\epsfysize=20cm \epsfbox{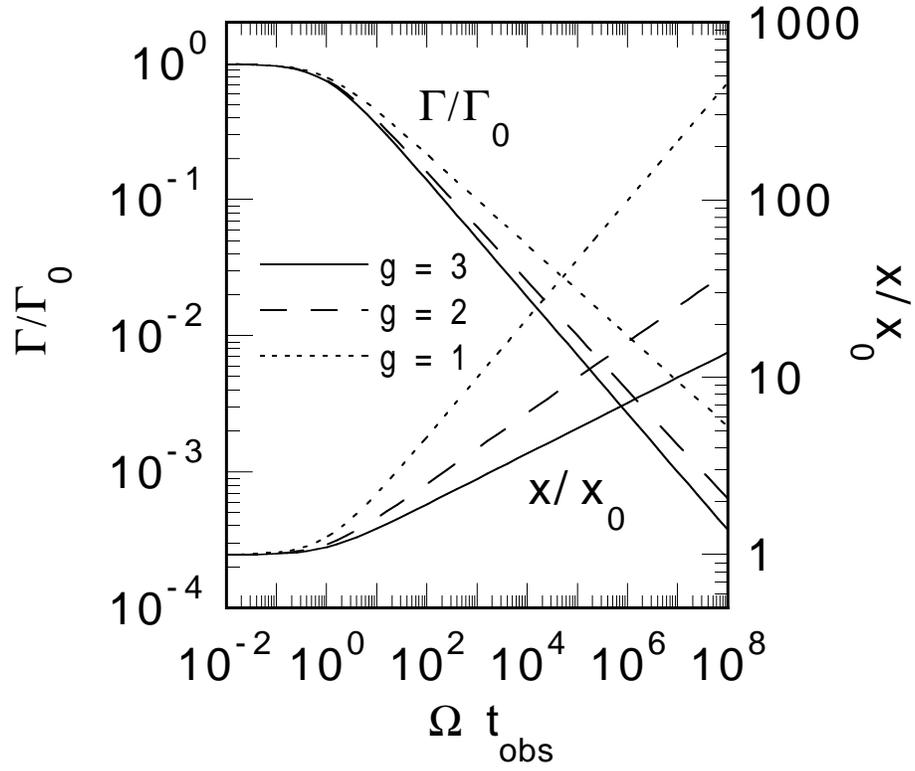}}
\vspace{-1.0in}
\caption{Dependence of $\Gamma/\Gamma_0$ amd the ratio of distance traveled to distance $x_0$
given by equation (15) on the dimensionless quantity $\Omega \to$, where $\Omega$ is
given through equation (40) and $\to$ is the observer time.}
\end{figure}

\begin{figure}
\centerline{\epsfysize=15cm \epsfbox{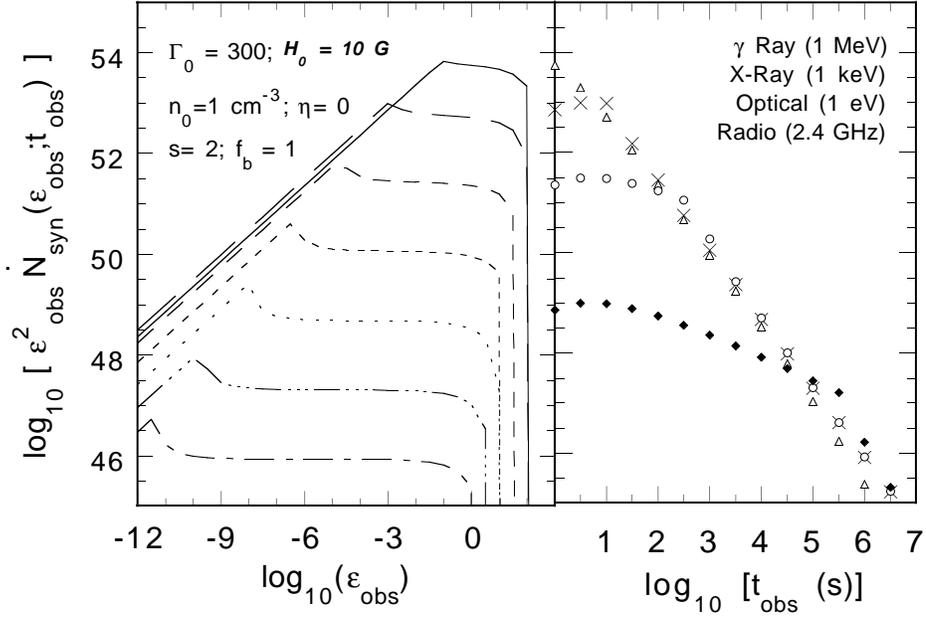}}
\vspace{-1.0in}
\caption{Spectra (left) and time profiles (right) of a blast wave decelerating in a uniform
external medium. Here it is assumed that the power-law behavior of the blast wave is
described by $g = g_r =3$, and that 100\% of the swept-up nonthermal proton energy is transformed
into an equally energetic power-law electron distribution with a low-energy cutoff $p_{\rm min} = 0.2$ and
a high-energy cutoff $p_{\rm max} = 3\times 10^6$. All other parameters are standard (see Table 1).
Observing times of the spectra are, from top to bottom, 1 sec, 10 sec, 100 sec, etc.  The time profiles
at 1 MeV, 1 keV, 1 eV, and 2.4 GHz are labeled by triangles, crosses, circles, and filled diamonds,
respectively.}
\end{figure}

\begin{figure}
\centerline{\epsfysize=15cm \epsfbox{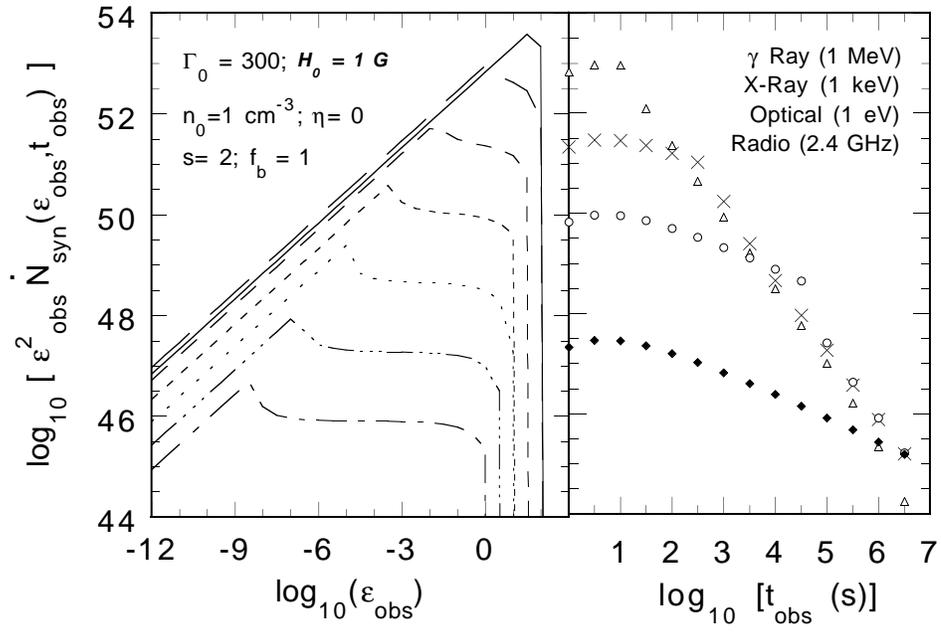}}
\vspace{-1.0in}
\caption{Same as Fig. 4, except that $H_0 = 1$ Gauss.}
\end{figure}

\begin{figure}
\centerline{\epsfysize=15cm \epsfbox{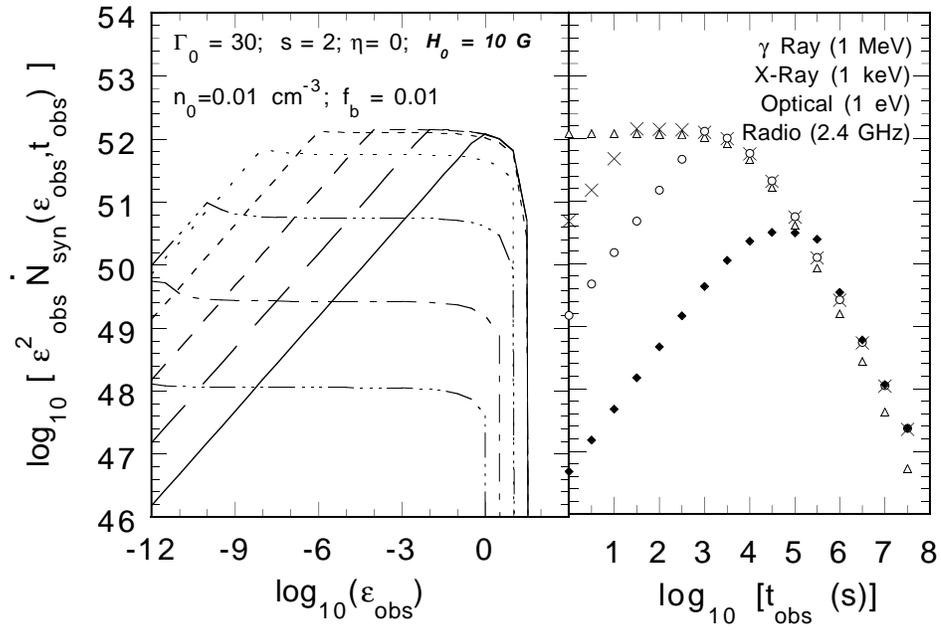}}
\vspace{-1.0in}
\caption{Same as Fig. 4, except that $\Gamma_0 = 30$, $n_0 = 0.01$ cm$^{-3}$, and $f_b = 0.01$.}
\end{figure}

\end{document}